\documentstyle[12pt]{article}
\newlength{\pubnumber} 

\catcode`\@=11
\@addtoreset{equation}{section}

\def\section{\@startsection{section}{1}{\z@}{3.5ex plus 1ex minus .2ex}
 {2.3ex plus .2ex}{\large\bf}}
\def\subsection{\@startsection{subsection}{2}{\z@}{2.3ex plus .2ex}
 {2.3ex plus .2ex}{\bf}}

\begin{document}

\begin{titlepage}
\samepage{
\setcounter{page}{1}
\rightline{IASSNS-HEP-97/67}
\rightline{\tt hep-th/9707148}
\rightline{July 1997}
\vfill
\begin{center}
   {\Large \bf Duality without Supersymmetry:\\
       The Case of the $SO(16)\times SO(16)$ String\\}
\vfill
   {\large
    Julie D. Blum\footnote{
     E-mail address: julie@sns.ias.edu}
     $\,$and$\,$
      Keith R. Dienes\footnote{
     E-mail address: dienes@sns.ias.edu}
    \\}
\vspace{.12in}
 {\it  School of Natural Sciences, Institute for Advanced Study\\
  Olden Lane, Princeton, N.J.~~08540~ USA\\}
\end{center}
\vfill
\begin{abstract}
  {\rm
   We extend strong/weak coupling duality to
   string theories without spacetime supersymmetry,
   and focus on the case of the unique ten-dimensional,
   nonsupersymmetric, tachyon-free $SO(16)\times SO(16)$ heterotic string.
   We construct a tachyon-free heterotic string model that interpolates
smoothly
   between this string and the ten-dimensional supersymmetric $SO(32)$
   heterotic string, and we construct a dual for this interpolating model.
   We find that the perturbative massless states of our dual theories precisely
   match within a certain range of the interpolation.
   Further evidence for this proposed duality
   comes from a calculation of the one-loop cosmological constant in both
theories,
   as well as the presence of a soliton in the dual theory.
   This is therefore the first known duality relation
   between nonsupersymmetric tachyon-free string theories.  Using this duality, we
   then investigate the perturbative and nonperturbative stability of the
   $SO(16)\times SO(16)$ string, and present a conjecture concerning its
ultimate
   fate.
   }
\end{abstract}
\vfill
\smallskip}
\end{titlepage}

\setcounter{footnote}{0}

\def\beq{\begin{equation}}
\def\eeq{\end{equation}}
\def\beqn{\begin{eqnarray}}
\def\eeqn{\end{eqnarray}}
\def\sosixteen{{$SO(16)\times SO(16)$}}
\def\V#1{{\bf V_{#1}}}
\def\half{{\textstyle{1\over 2}}}
\def\ttwo{{\vartheta_2}}
\def\tthree{{\vartheta_3}}
\def\tfour{{\vartheta_4}}
\def\ttwob{{\overline{\vartheta}_2}}
\def\tthreeb{{\overline{\vartheta}_3}}
\def\tfourb{{\overline{\vartheta}_4}}
\def\etainv{{\overline{\eta}}}
\def\Str{{{\rm Str}\,}}
\def\bone{{\bf 1}}
\def\chibar{{\overline{\chi}}}
\def\Jbar{{\overline{J}}}
\def\qbar{{\overline{q}}}
\def\calO{{\cal O}}
\def\calE{{\cal E}}
\def\calT{{\cal T}}
\def\calM{{\cal M}}
\def\calF{{\cal F}}
\def\calY{{\cal Y}}
\def\rep#1{{\bf {#1}}}
\def\tenn{{(10)}}
\def\ninen{{(9)}}
\hyphenation{su-per-sym-met-ric non-su-per-sym-met-ric}
\hyphenation{space-time-super-sym-met-ric}
\hyphenation{mod-u-lar mod-u-lar--in-var-i-ant}


\def\inbar{\,\vrule height1.5ex width.4pt depth0pt}

\def\IC{\relax\hbox{$\inbar\kern-.3em{\rm C}$}}
\def\IQ{\relax\hbox{$\inbar\kern-.3em{\rm Q}$}}
\def\IR{\relax{\rm I\kern-.18em R}}
 \font\cmss=cmss10 \font\cmsss=cmss10 at 7pt
\def\IZ{\relax\ifmmode\mathchoice
 {\hbox{\cmss Z\kern-.4em Z}}{\hbox{\cmss Z\kern-.4em Z}}
 {\lower.9pt\hbox{\cmsss Z\kern-.4em Z}}
 {\lower1.2pt\hbox{\cmsss Z\kern-.4em Z}}\else{\cmss Z\kern-.4em Z}\fi}

\def\NPB#1#2#3{{\it Nucl.\ Phys.}\/ {\bf B#1} (19#2) #3}
\def\PLB#1#2#3{{\it Phys.\ Lett.}\/ {\bf B#1} (19#2) #3}
\def\PRD#1#2#3{{\it Phys.\ Rev.}\/ {\bf D#1} (19#2) #3}
\def\PRL#1#2#3{{\it Phys.\ Rev.\ Lett.}\/ {\bf #1} (19#2) #3}
\def\PRT#1#2#3{{\it Phys.\ Rep.}\/ {\bf#1} (19#2) #3}
\def\CMP#1#2#3{{\it Commun.\ Math.\ Phys.}\/ {\bf#1} (19#2) #3}
\def\MODA#1#2#3{{\it Mod.\ Phys.\ Lett.}\/ {\bf A#1} (19#2) #3}
\def\IJMP#1#2#3{{\it Int.\ J.\ Mod.\ Phys.}\/ {\bf A#1} (19#2) #3}
\def\NUVC#1#2#3{{\it Nuovo Cimento}\/ {\bf #1A} (#2) #3}
\def\etal{{\it et al.\/}}

\long\def\@caption#1[#2]#3{\par\addcontentsline{\csname
  ext@#1\endcsname}{#1}{\protect\numberline{\csname
  the#1\endcsname}{\ignorespaces #2}}\begingroup
    \small
    \@parboxrestore
    \@makecaption{\csname fnum@#1\endcsname}{\ignorespaces #3}\par
  \endgroup}
\catcode`@=12

\input epsf

\section{Introduction}
\setcounter{footnote}{0}

Much has been learned about the nonperturbative properties of superstring
theories by mapping these theories at strong coupling into other theories
at weak coupling.  The constraints of supersymmetry provide much of
the structure of these duality relations, but one would like
to understand whether duality might be a more general property of
string theory which holds even without supersymmetry.
One might then begin to understand the nonperturbative properties of
nonsupersymmetric theories as well.

As a step in this direction, we will study nonsupersymmetric duality
by smoothly connecting nonsupersymmetric ten-dimensional theories
to supersymmetric ten-dimensional theories.
Such connections can be achieved in nine dimensions through {\it interpolating
models}.
Dimensional reduction as a means of breaking supersymmetry in string theory
was originally
proposed in Ref.~\cite{one}, and interpolating models of this sort
have been discussed in Refs.~\cite{one,three,two}.

As our focus, we shall consider the case of the \sosixteen\ string \cite{four}.
This is the unique, tachyon-free, nonsupersymmetric heterotic string theory in
ten dimensions, and it can therefore be viewed as the nonsupersymmetric
counterpart
of the supersymmetric $SO(32)$ and $E_8\times E_8$ strings.
The absence of tachyons in this theory is very important, implying
that its one-loop cosmological constant is finite.
This makes a stability analysis possible.  The absence of tachyons also
guarantees a number of other remarkable properties, such as the existence
of a hidden ``misaligned supersymmetry'' \cite{misaligned} in the string
spectrum.

As a means of investigating strong/weak duality relations for the \sosixteen\
string,
we will use the interpolation idea and begin by constructing a heterotic
interpolating model
that smoothly connects the nonsupersymmetric $SO(16)\times SO(16)$ string to
the supersymmetric $SO(32)$ string.
Our model will be a twisted compactification of the supersymmetric $SO(32)$
heterotic string
on a circle of radius $R_H$.  Our twists will be chosen in such a way
that the theory reproduces the ten-dimensional supersymmetric $SO(32)$ theory
as $R_H\to \infty$, but is related by $T$-duality to the
ten-dimensional nonsupersymmetric \sosixteen\ string as $R_H\to 0$.

Our next step will be to construct the strong-coupling dual of this
interpolating
model. The fact that this interpolating model continuously connects the
\sosixteen\ theory
to the supersymmetric $SO(32)$ theory suggests that the dual
of our interpolating model is an open-string  model.
We therefore study the possible interpolations of open-string models.
We find that we are able to construct a corresponding nine-dimensional
open-string model which interpolates between the ten-dimensional $SO(32)$
Type~I theory
as $R_I\to \infty$,
and a ten-dimensional nonsupersymmetric, tachyonic open-string
theory at $R_I=0$.\footnote{
     This tachyonic theory is equivalent to
     the $R_I=0$ compactification of one of the ten-dimensional tachyonic
     open-string theories originally constructed in Ref.~\cite{Sagnotti}.
     This ten-dimensional theory has recently been conjectured \cite{BG} to
     be dual to the bosonic string compactified to ten dimensions.}
It might initially seem that the appearance of tachyons in our open-string
interpolating model might ruin our attempted duality relation, or at the very
least render it meaningless.  However, there exists a critical radius $R^\ast$
above which
our open-string interpolating model is completely free of tachyons, and it is
in
precisely this range that we find evidence for an exact duality relation.
Specifically, we find that for $R_I>R^{\ast}$,
the perturbative massless spectrum of the open-string interpolating model
agrees
exactly with that of our corresponding heterotic interpolating model.
Furthermore, our open-string model contains a soliton which (modulo several
assumptions)
behaves as a fundamental $SO(16)\times SO(16)$ nonsupersymmetric heterotic
string when the ten-dimensional Type~I coupling $\lambda^{(10)}_I$ is
large, or equivalently when the ten-dimensional heterotic coupling
$\lambda^{(10)}_H=1/\lambda^{(10)}_I$ is small.
Thus, if our interpretation is correct,
this duality between our heterotic and open-string interpolating models
would be the first known example of a strong/weak coupling duality
relation between nonsupersymmetric, tachyon-free string theories.

Our results for this model --- as well as similar results
for several other nonsupersymmetric interpolating models ---
will be given in great detail in Ref.~\cite{BD}.
Here we shall merely outline and summarize our results
for the $SO(16)\times SO(16)$ case.  In Sect.~2, we describe
our heterotic interpolating model and discuss its
one-loop cosmological constant.  Then, in Sect.~3, we construct our
proposed dual model, and discuss its
one-loop cosmological constant as well.  In Sect.~4, we examine
the soliton of our dual open-string theory, and in Sect.~5
we make use of our duality relation in order to
analyze the stability of these models and make a conjecture about the
behavior of the $SO(16)\times SO(16)$
nonsupersymmetric heterotic string.
We find that we are able to propose a phase diagram that describes
the behavior of this string as a function of its coupling and radius.
Finally, Sect.~6 contains our conclusions.

\section{The \sosixteen\ Heterotic Interpolating Model}
\setcounter{footnote}{0}

Our goal in this section is to construct a model which interpolates
between the ten-dimensional $SO(32)$
supersymmetric heterotic string
and the ten-dimensional $SO(16)\times SO(16)$ nonsupersymmetric heterotic
string.
To do this, we compactify the $SO(32)$ theory on a
circle of radius $R_H$ and orbifold by a $\IZ_2$ element $Q'_H\equiv{\cal T}
Q_H$ defined as follows.  The operator $\cal T$ is a translation of the
coordinate
$x_1$ of the circle by half of the circumference of the circle:
\beq
       {{\cal T}:\hskip .2cm x_1\rightarrow x_1+\pi R_H ~.}
\label{T}
\eeq
The operator $Q_H$ is the generator of the
orbifold
that produces the ten-dimensional nonsupersymmetric $SO(16)\times SO(16)$
theory
from the ten-dimensional supersymmetric $SO(32)$ theory.
Note that if we decompose the representations of the original gauge group
$SO(32)$
into those of $SO(16)\times SO(16)$,
then $Q_H$ can be written as $(-1)^F {\cal R}$ where
$F$ is the spacetime fermion number and where ${\cal R}$ acts with a minus sign
on both of the spinors of one of the $SO(16)$ factors, and
with a minus sign on the vector and one of the spinor
representations of the other factor.

Projection by
the half-rotation operator $\calT$ by itself has the effect of halving the
radius of
the circle, for odd momentum states on the circle are not invariant under
$\calT$
and are projected out of the spectrum.
However, at infinite radius,
the odd momentum states are
degenerate with the even momentum states.
This causes $\calT$ to act as zero in this case,
which merely amounts to a rescaling of the effective volume normalization of
the $SO(32)$ partition function.
We thus recover the ten-dimensional supersymmetric $SO(32)$ theory.
At zero radius, all momentum states are infinitely massive so
the orbifold action reduces to $Q_H$.  This gives us the
$SO(16)\times SO(16)$ theory at $R_H=0$, and by $T$-duality this
theory is equivalent to the uncompactified $SO(16)\times SO(16)$ theory in ten
dimensions.
Spacetime supersymmetry is broken for all finite radii, and there
are no tachyons in this model for any radius.  Note
also that the $SO(32)$ gauge symmetry is restored only at the infinite-radius
endpoint,
and there are no enhanced gauge symmetries at any finite radius.
For generic radii in the range $0<R_H<\infty$, the massless states of this model 
consist of the graviton, antisymmetric tensor, dilaton, two $U(1)$ vectors, 
vectors in the adjoint representation of $SO(16)\times SO(16)$, and a spinor in the
$(\rep{16},\rep{16})$
representation.  
At the discrete radius $R_H=\sqrt{2\alpha'}$ there are also two massless scalars 
in the $(\rep{1},\rep{128})\oplus (\rep{128},\rep{1})$ representation,
and at the discrete radius $R_H=\sqrt{\alpha'}$ there are 
two massless spinors that are singlets under $SO(16)\times SO(16)$. 
As $R_H\to 0$, a massive  opposite-chirality spinor in the
$(\rep{1},\rep{128}) \oplus (\rep{128},\rep{1})$ representation
becomes massless, and the $U(1)$ gauge bosons disappear.

\begin{figure}[tb]
\centerline{
         \epsfxsize 2.9 truein \epsfbox {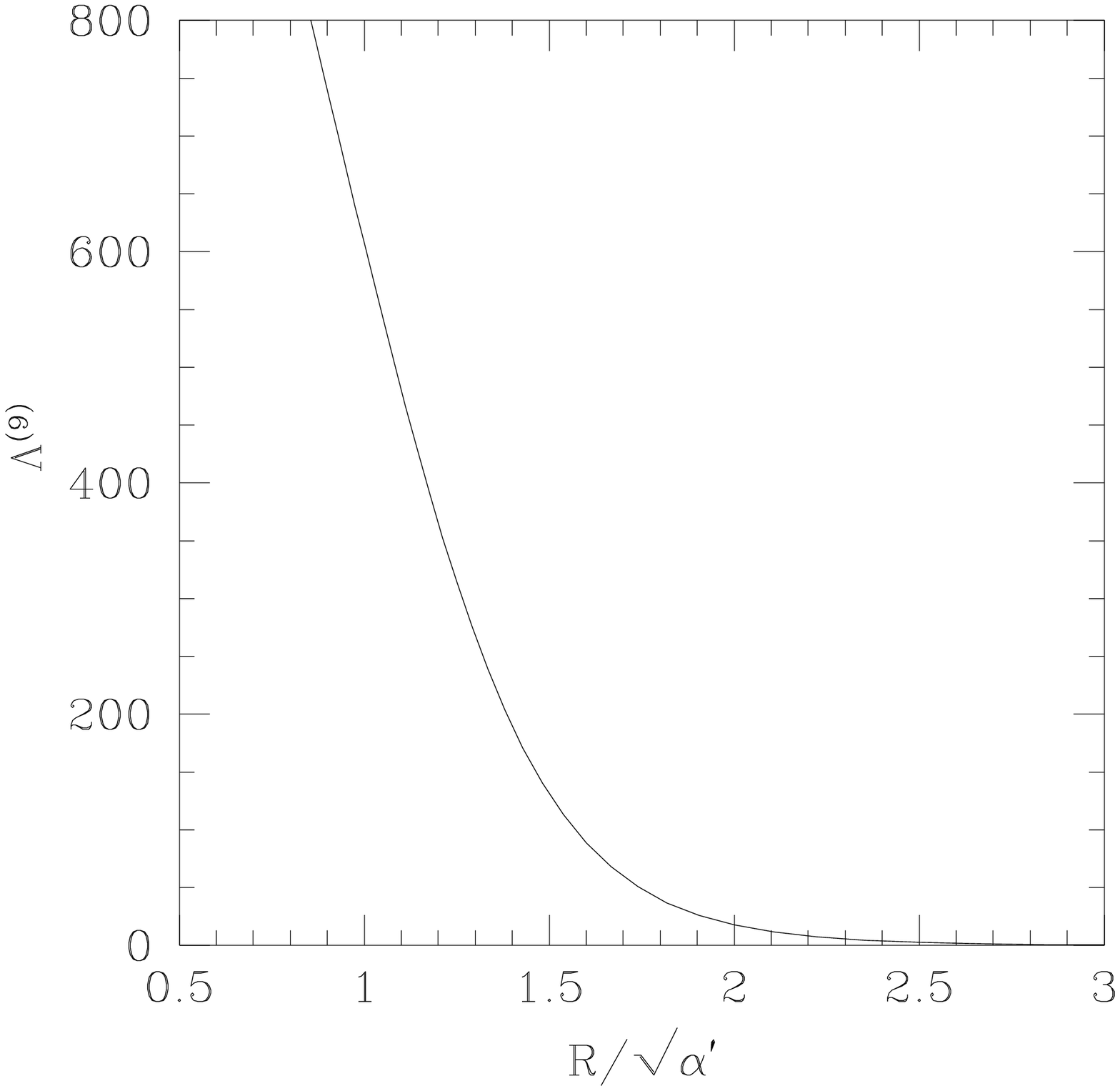}
         \hskip 0.2 truein
         \epsfxsize 2.9 truein \epsfbox {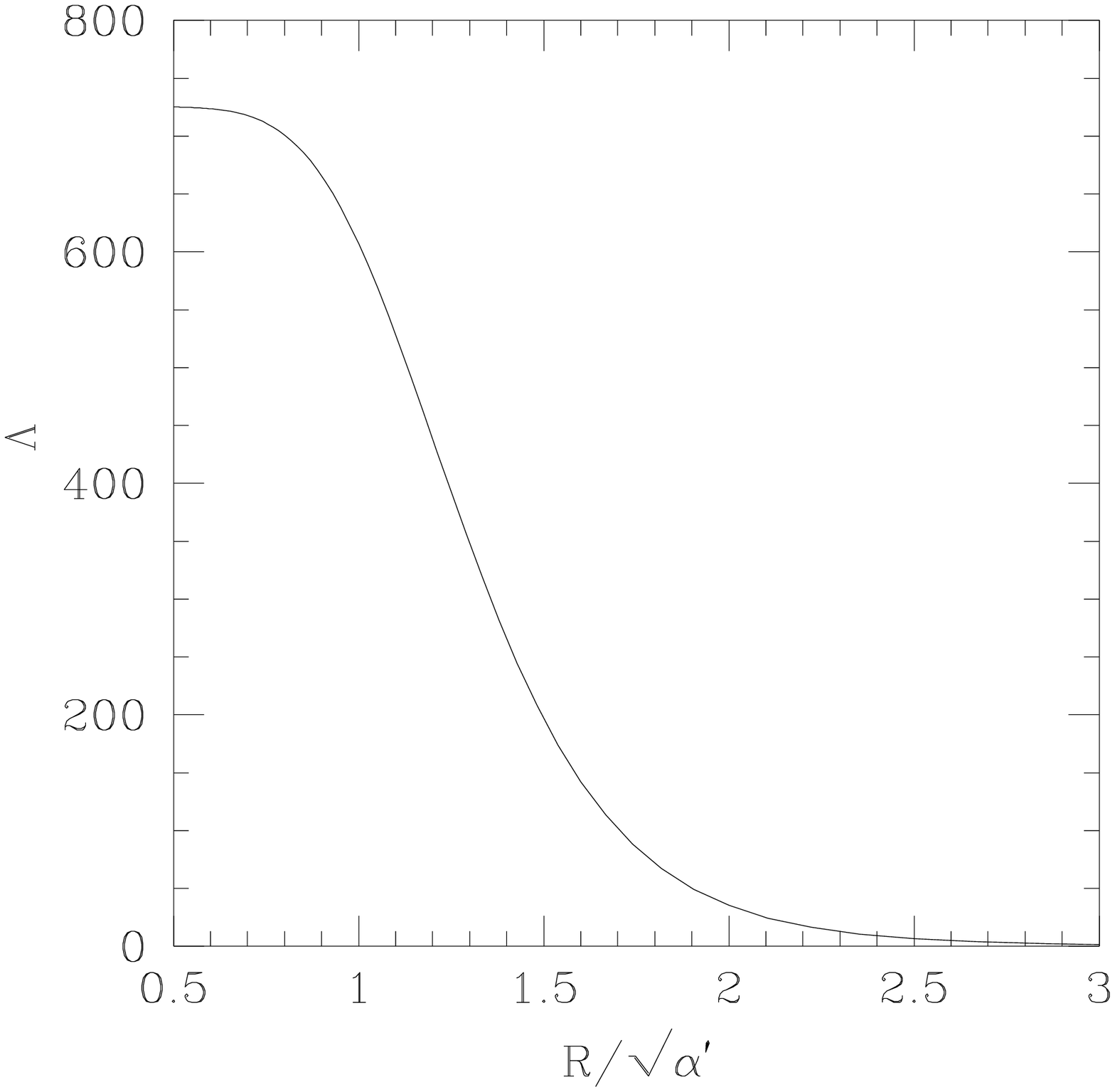}
  }
\caption{
  The one-loop cosmological constants
    $\Lambda^{(9)}$ (left plot) and $\tilde \Lambda$ (right plot)
     as functions of the radius $R_H$ of the compactified dimension,
     in units of $\half\calM^{9}$ and $\half\calM^{10}$ respectively.
   This model reproduces the supersymmetric $SO(32)$ heterotic string
    as $R_H\to \infty$ and the nonsupersymmetric \sosixteen\ heterotic
    string as $R_H\to 0$.}
\label{interplambda}
\end{figure}

The one-loop vacuum amplitude or cosmological constant of this model
can be calculated, and the results are shown in Fig.~\ref{interplambda}.
Note that we have taken
\beq
     \Lambda^{(9)}(R_H) ~\equiv~  -\,\half \,\calM^9\,
            \int_{\cal F} {d^2 \tau \over ({\rm Im}\, \tau)^2} \,Z(\tau,R_H)
\eeq
where $Z(\tau,R_H)$ is the radius-dependent partition function of the
interpolating
model and where $\calM\equiv (4\pi^2 \alpha')^{-1/2}$.  We can also define
\beq
     \tilde \Lambda (R_H) ~\equiv~ \calM\, {R_H\over \sqrt{\alpha'}}\,
\Lambda^{(9)}(R_H)~.
\label{tildelambdadef}
\eeq
Thus $\lim_{R_H\to 0} \tilde\Lambda(R_H)$ is finite, and gives the
ten-dimensional
cosmological constant of the \sosixteen\ string.

Several observations are worth mentioning.  The nine-dimensional
cosmological constant $\Lambda^\ninen$ is
positive and finite for all finite, nonzero radii.  At infinite radius,
the cosmological constant vanishes, and there is a critical point with zero
slope.
The slope of the curve is negative for all finite positive radii.

\section{The Dual Interpolating Model}
\setcounter{footnote}{0}

We now construct our proposed dual for the \sosixteen\
heterotic interpolating model of Sect.~2.

Our method for constructing open-string interpolating models will be as
follows.
We take the point of view that a given open-string model is consistent if
it can be realized as the orientifold \cite{orientifolds,Sagnotti}
of a Type~II model, and if it
satisfies tadpole anomaly-cancellation constraints.
We therefore begin by examining Type~II interpolating models,
and in particular we construct a
nine-dimensional interpolating model
that connects the supersymmetric Type~IIB theory at infinite radius
to the nonsupersymmetric so-called Type~0B theory
\cite{five} at zero radius.  At zero radius,
the nonsupersymmetric Type~0B endpoint is equivalent
by $T$-duality to the so-called Type~0A theory \cite{five}.
This Type~II interpolating model is similar to that considered in
Ref.~\cite{six}.
In particular, it can be realized by compactifying the Type~IIB theory
on a circle of radius $R_I$, and then orbifolding by the action
$g=\calT (-1)^F$.  Here $\calT$ is again the operator given in Eq.~(\ref{T}),
and $F\equiv F_L+F_R$ is the total Type~II spacetime fermion number.
Because of the factor $(-1)^F$ within $g$,
orbifolding by $g$ breaks supersymmetry for all radii $R_I<\infty$.

Given this Type~II interpolating model, we then
construct its corresponding open-string orientifold.  We take
our orientifold generator to be $\Omega$, the worldsheet parity operator.
Thus, putting our entire construction together, we see that our open-string
interpolating model is
a $\IZ_2\times \IZ_2$ orientifold of the supersymmetric Type~IIB
theory compactified on a circle of radius $R_I$.
The first $\IZ_2$ factor corresponds to the worldsheet parity operator
$\Omega$,
and the second to the orbifold operator $g$.
Because these two operators commute, we can consider our interpolating
open-string model either as an orientifold of the Type~II interpolation,
or as an orbifold of the $SO(32)$ Type~I theory.
As $R_I\to\infty$, this open-string model smoothly reproduces the
supersymmetric
$SO(32)$ Type~I theory.
Further details concerning the construction of this
model can be found in Ref.~\cite{BD}.

The one-loop amplitudes for our orientifold model can be calculated for the
torus, Klein bottle, cylinder, and M\"obius strip.
As is usual for orientifolds, for each of the elements
$\Omega$ and $g$  we must specify an action
on the Chan-Paton factors of the open-string sectors.
Remarkably, it then turns out \cite{BD} that
both the NS-NS and Ramond-Ramond massless tadpole divergences
will be cancelled for $R_I>0$ if the gauge group is \sosixteen.
This choice also avoids introducing
open-string tachyonic divergences.

This open-string interpolating model contains a critical
radius $R^\ast\equiv 2\sqrt{2\alpha'}$.
For radii $R_I$ in the range $R^{\ast}<R_I<\infty$, the massless spectrum
of our model consists of the graviton, antisymmetric tensor, dilaton,
two $U(1)$ vectors, vectors in the adjoint of $SO(16)\times SO(16)$,
and a spinor in the $(\rep{16},\rep{16})$ representation.
There are no tachyons in either the closed-string or open-string sectors.
The massless spectrum of this model
therefore agrees precisely with that of its heterotic dual in this range.
We emphasize
that a perturbative comparison of the spectra of these dual theories is
only possible in the nontachyonic range of the interpolation with $R_I>R^\ast$.

At $R_I=R^{\ast}=2 \sqrt{ 2 \alpha'}$, two
additional massless closed-string states appear,
and these become  tachyonic for $R_I<R^{\ast}$.
In fact, an infinite number of closed-string tachyons appear at discrete radii
in the range
$R_I<R^{\ast}$.  Nevertheless, all tadpole anomalies continue to cancel
for $R_I>0$.

\begin{figure}[tb]
\centerline{\epsfxsize 4.0 truein \epsfbox {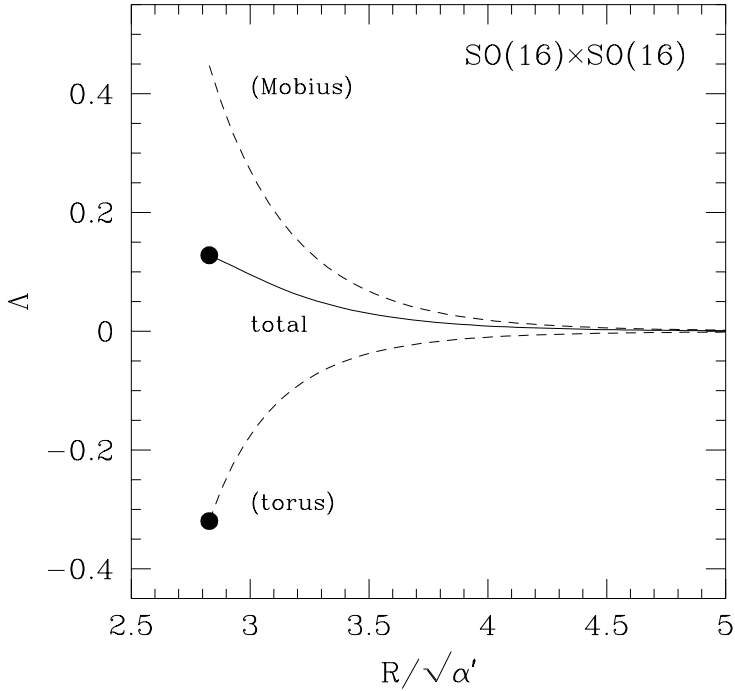}}
\caption{
  The total one-loop cosmological constant
    $\tilde \Lambda$
   for our open-string interpolating model, plotted in units of $\half
\calM^{10}$,
   as a function of the radius $R_I$ of the compactified dimension.
   Also shown (dashed lines) are the separate torus and M\"obius-strip
contributions;
   note that the Klein-bottle and cylinder contributions vanish.
   This open-string model reproduces the supersymmetric $SO(32)$ Type~I string
    as $R_I\to \infty$, and has gauge group $SO(16)\times SO(16)$ for all
finite
    radii.  The torus amplitude develops
    a divergence below $R^\ast/\sqrt{\alpha'}\equiv 2\sqrt{2}\approx 2.83$,
    which reflects the appearance of a tachyon in the
    torus amplitude below this radius.  All other contributions are
     tachyon-free for all radii.}
\label{interplambdaI}
\end{figure}

Just as for our heterotic interpolating model,
we can calculate the one-loop cosmological constant for our open-string model.
 {\it A priori}\/, this receives four contributions, one each from
the torus, Klein bottle, cylinder, and M\"obius strip.
However, for this model, the Klein bottle and cylinder contributions
vanish exactly.
This occurs because the Klein bottle is immune to supersymmetry breaking
for $R_I>0$, and because we have chosen the gauge symmetry to be \sosixteen.
Thus only the torus and M\"obius contributions remain, and
these contributions are shown in Fig.~\ref{interplambdaI}.
Note that for this figure, we have used the alternate definition
\beq
     \tilde \Lambda (R_I) ~\equiv~ \calM\, {\sqrt{\alpha'}\over R_I}\,
\Lambda^{(9)}(R_I)~
\label{tildelambdadefI}
\eeq
where $\Lambda^{(9)}$ is defined as the (appropriately weighted) sum of the
modular integrals of the torus and M\"obius strip partition functions.
This definition ensures that $\tilde \Lambda(R_I)$ formally reproduces
the Type~I $SO(32)$ cosmological constant as $R_I\to \infty$.

It is clear from this figure that
the torus contribution to the cosmological constant
is strictly negative.  Fortunately,
the M\"obius contribution more than cancels the torus
contribution, and produces a net positive cosmological constant for all radii
$R^{\ast}<R_I<\infty$.
It turns out that
$SO(16)\times SO(16)$ is the unique gauge symmetry
that can be chosen in our open-string model for which the net cosmological
constant
is positive rather than negative for $R_I>R^{\ast}$.
Moreover, the slope of the curve is strictly negative in
this range, and vanishes at infinity.

It is interesting that
the inclusion of open-string sectors in tachyon-free theories
can provide competing contributions to the cosmological constant.
This suggests that it might be possible to
cancel the contributions of open-string sectors against those of
closed-string sectors in order to provide a new mechanism for cancelling
the cosmological constant in open-string theories.

As we have remarked above,
closed-string tachyons appear in this model at $R_I<R^{\ast}$.
There is therefore a discontinuity
in the cosmological constant at the critical point $R_I=R^\ast$,
suggestive of a phase transition in our open-string theory.
We shall explore this idea further in Sect.~5.

\section{The Soliton}
\setcounter{footnote}{0}

Our open-string model contains a solitonic object or D1-brane.
Solitons of the $SO(32)$ Type~I theory corresponding to $SO(32)$ heterotic
strings
were found as classical solutions in Ref.~\cite{seven}, and were constructed
from
a collective-coordinate expansion on the D1-brane in Ref.~\cite{eight}.
In this section, we will construct
the soliton of our model by projecting
the worldsheet fields of the $SO(32)$ soliton
by the orbifold element $g$.
This procedure involves numerous subtleties, and a fuller discussion
can be found in Ref.~\cite{BD}.
Here we summarize only the salient features.

The tension of the soliton is $T_S\sim T_F/\lambda_I^{(10)}$
where $T_F=1/2\pi\alpha'$
is the fundamental string tension and where $\lambda_I^{(10)}$ is the
ten-dimensional Type~I coupling.
If we were to choose the soliton to lie along any direction
orthogonal to the compact direction, the soliton would be very heavy for
all perturbative values of the coupling and would not behave as a fundamental
object.  Thus, we shall choose the soliton to lie along the compact $x_1$
direction because this choice allows states of the soliton to become nearly
massless
at perturbative values of the couplings for sufficiently small $R_I$.
Note, though, that our soliton is not expected to behave fundamentally in the
range of the interpolation where a perturbative analysis is valid.

In quantizing the fields on the soliton, we find that the quantized
zero-mode momentum of Type~I bosons in the $x_1$ direction becomes the
quantized oscillator moding of the worldsheet fields on the soliton.
Moreover, the Type~I theory
provides a restriction on the left- and right-moving
momenta of the form  $p_1^L=p_1^R$, and this restriction becomes the
worldsheet restriction $L_0=\overline{L_0}$
on the left- and right-moving Virasoro generators on the soliton.

We will project by the $\IZ_2$ element $\calY={\cal T}(-1)^F\gamma_g$ where
$\gamma_g$ is the Type~I Wilson line that breaks $SO(32)$ to $SO(16)\times
SO(16)$.
It turns out that the action of $(-1)^F\gamma_g$ on the worldsheet
fields of the $SO(32)$ soliton exactly corresponds to $Q_H$.  Since
$(-1)^F \gamma_g$ does not act on the Type~I spacetime, the spacetime action of
$\calY$
is solely due to $\cal T$, and  effectively halves the radius of the circle.
All worldsheet bosons of the soliton will thus be restricted to
integral moding such that these fields are periodic on the half-radius circle.

Before the projection, the right-movers along $x_1$ consist of eight transverse
spatial bosonic fields $X^i$, $i=1,...,8$, and a Green-Schwarz fermion $S^a$,
$a=1,...,8$,
of definite chirality.
The left-movers consist of eight transverse spatial bosonic fields $X^i$
and 32 worldsheet fermions $\psi^A$, $A=1,...,32$.  There is also
a GSO projection acting on the left-moving fermions.  It is convenient in our
analysis to
separate the fields on the half-radius circle into two groups.  The ``Class~A''
sectors will
be those for which all fields have the usual modings on the half-radius circle
(such that the fields are periodic on the half-radius circle),
while the ``Class~B'' sectors will be those in which some fields may be
antiperiodic on the half-radius circle (provided they are consistent with the
symmetries preserved by the orbifold).  This describes the Ramond sector of the
soliton, and just as for the supersymmetric $SO(32)$ soliton, we will assume
that a Neveu-Schwarz sector arises nonperturbatively.

We emphasize that it is the projection onto the half-radius circle
that gives rise to the Class~B sectors.
Specifically,
we have {\it not}\/ added the Class~B sectors by hand --- they  instead appear
naturally when the original fields on the circle of radius $R_I$ are
decomposed into fields on the circle of half-radius $R_I/2$.
This passing from the original radius to the half-radius naturally generates
the unexpected modings.
Note that because of their nontraditional modings,
all Class~B sectors would be ordinarily be projected out of the spectrum
by the operator $\calT$ alone.

We can summarize the resulting sectors in the following tables where the
identity,
vector, spinor, and conjugate-spinor representations of the right-moving
transverse
Lorentz group $SO(8)$ and the
two left-moving $SO(16)$ gauge factors are labelled in an obvious way.  For
each, we have also
indicated the corresponding eigenvalue under $\calY$.
The Class~A sectors are:
\beq
\begin{tabular}{||c|c||c|c||c|c||}
\hline
  \multicolumn{2}{||c||}{right-movers} & \multicolumn{4}{c||}{left-movers}
\\
\hline
\hline
  \multicolumn{2}{||c||}{~} & \multicolumn{2}{c||}{Ramond} &
\multicolumn{2}{c||}{NS} \\
\hline
   $\calY$ & sector & $\calY$ & sector &  $\calY$ & sector \\
\hline
 $+1$ & $V_8$ & $+1$ & $S_{16}^{(1)}  S_{16}^{(2)}$
    &  $+1$ & $I_{16}^{(1)}  I_{16}^{(2)}$ \\
 $-1$ & $S_8$ & $-1$ & $C_{16}^{(1)}  C_{16}^{(2)}$
    & $-1$ & $V_{16}^{(1)}  V_{16}^{(2)}$ \\
\hline
\end{tabular}
\label{classAtable}
\eeq
and the Class~B sectors are:
\beq
\begin{tabular}{||c|c||c|c||c|c||}
\hline
  \multicolumn{2}{||c||}{right-movers} & \multicolumn{4}{c||}{left-movers}
\\
\hline
\hline
  \multicolumn{2}{||c||}{~} &
  \multicolumn{2}{c||}{Ramond} &
  \multicolumn{2}{c||}{NS}  \\
\hline
   $\calY$ & sector & $\calY$ & sector & $\calY$ & sector \\
\hline
 $+1$ & $I_8$
      & $+1$ & $V_{16}^{(1)}  I_{16}^{(2)}$
           & $+1$ & $S_{16}^{(1)}  C_{16}^{(2)}$ \\
 $+1$ & $C_8$
      & $-1$ & $I_{16}^{(1)}  V_{16}^{(2)}$
           & $-1$ & $C_{16}^{(1)}  S_{16}^{(2)}$ \\
    ~ & ~
      & $+1$ & $I_{16}^{(1)}  S_{16}^{(2)}$
           & $+1$ & $S_{16}^{(1)}  I_{16}^{(2)}$ \\
    ~ & ~
      & $+1$ & $V_{16}^{(1)}  C_{16}^{(2)}$
           & $+1$ & $C_{16}^{(1)}  V_{16}^{(2)}$ \\
    ~ & ~
      & $-1$ & $C_{16}^{(1)}  I_{16}^{(2)}$
           & $+1$ & $I_{16}^{(1)}  C_{16}^{(2)}$ \\
    ~ & ~
      & $-1$ & $S_{16}^{(1)}  V_{16}^{(2)}$
           & $+1$ & $V_{16}^{(1)}  S_{16}^{(2)}$ \\
\hline
\end{tabular}
\label{classBtable}
\eeq

The final step, then, is to join these left- and right-moving sectors together.
In order to do this, we make the following assumptions.
First, as discussed above, we impose the requirement that $L_0=\overline{L_0}$.
Second,  we require that the exchange
symmetry of the two $SO(16)$ gauge factors be preserved.
Third, we require that all left/right combinations
be invariant under $\calY$.
And finally, we demand that the Class~A and Class~B not be permitted to mix
when joining left- and right-moving sectors.
If we impose all four of these restrictions simultaneously,
we obtain precisely the sectors that comprise the
nonsupersymmetric $SO(16)\times SO(16)$ heterotic string:
\beqn
   {\rm Class~A:}&&~~~
     V_8 \,  I_{16}^{(1)} I_{16}^{(2)} ~,~~
     V_8 \, S_{16}^{(1)} S_{16}^{(2)} ~,~~
     S_8 \, V_{16}^{(1)} V_{16}^{(2)} ~,~~
     S_8 \, C_{16}^{(1)} C_{16}^{(2)}  \nonumber\\
   {\rm Class~B:}&&~~~
     I_8 \,  V_{16}^{(1)} C_{16}^{(2)} ~,~~
     I_8 \,  C_{16}^{(1)} V_{16}^{(2)} ~,~~
     C_8 \,  I_{16}^{(1)} S_{16}^{(2)} ~,~~
     C_8 \,  S_{16}^{(1)} I_{16}^{(2)} ~.
\label{resultingsectors}
\eeqn

It is apparent
from this correspondence between our soliton and the $SO(16)\times SO(16)$
nonsupersymmetric string that the Class~A sectors correspond to
the untwisted sectors of the heterotic theory while the Class~B sectors
correspond to
the twisted sectors of the heterotic theory.
We find it quite miraculous that these twisted sectors, which are required
by modular invariance on the heterotic side, automatically appear on the Type~I
side via the above simple decomposition and projection by $\calY$.
This indicates that the D1-brane soliton
somehow manages to reconstruct modular invariance, even without supersymmetry.
This issue will be discussed more fully in Ref.~\cite{BD}.

Note that our third restriction permits combinations of left- and right-moving
sectors that are not separately invariant under $\calY$.  This can
be interpreted as an interaction between such sectors, and makes sense
at strong coupling.  Likewise, our fourth restriction is essentially a
nonperturbative selection rule.
While both of these restrictions are plausible at strong coupling,
they are somewhat implausible at weak coupling.  However, it is very
interesting to
note that if we remove the fourth restriction
entirely and enforce a strict projection by $\calY$ on the left- and
right-moving
sectors separately,  then we instead obtain
the massless sectors of the $E_8\times E_8$ supersymmetric heterotic string:
\beqn
   &&~~~
     V_8 \,  I_{16}^{(1)} I_{16}^{(2)} ~,~~
     V_8 \, S_{16}^{(1)} S_{16}^{(2)} ~,~~
     V_8 \,  S_{16}^{(1)} I_{16}^{(2)} ~,~~
     V_8 \, I_{16}^{(1)} S_{16}^{(2)} ~,~~\nonumber\\
   &&~~~
     C_8 \,  I_{16}^{(1)} I_{16}^{(2)} ~,~~
     C_8 \, S_{16}^{(1)} S_{16}^{(2)} ~,~~
     C_8 \,  S_{16}^{(1)} I_{16}^{(2)} ~,~~
     C_8 \, I_{16}^{(1)} S_{16}^{(2)} ~,~~\nonumber\\
   &&~~~
     I_8 \,  V_{16}^{(1)} C_{16}^{(2)} ~,~~
     I_8 \,  C_{16}^{(1)} V_{16}^{(2)} ~.~~
\eeqn
Indeed, the only extra sectors in the above list that are not found in the 
$E_8\times E_8$ string are those in the last line above.
These are purely massive, however, and become infinitely massive
as $R_I\rightarrow 0$.  They can therefore be expected to decouple completely.

We thus conclude that
at strong coupling, the soliton
is expected to behave as an \sosixteen\
heterotic string, whereas at weak coupling, the
massless worldsheet fields on the soliton
are those of an $E_8\times E_8$ heterotic string.

\section{The Fate of the $SO(16)\times SO(16)$ String:\\
  Analysis and Conjecture}
\setcounter{footnote}{0}

Using our candidate open-string dual, we now proceed to address questions of
stability.

At the supersymmetric $R_I=R_H=\infty$ endpoints, the relation
\beq
     {\lambda_I^{(10)}={1\over \lambda_H^{(10)}}}
\label{ws}
\eeq
is expected to be valid where $\lambda_H^{(10)}$
is the ten-dimensional heterotic coupling.  Since there is no perturbative
discontinuity in our orientifold for $R_I>R^{\ast}$, we will assume that
this relation continues to be valid there.

We will need to distinguish
the tachyon-free Phase I where $R_I>R^{\ast}$ and
$\lambda_H^{(10)}<R_H^2/{(R^{\ast})}^2$ from Phase II where $R_I<R^{\ast}$
and $\lambda_H^{(10)}>R_H^2/{(R^{\ast})}^2$.
For $R_I>R^{\ast}$, the massless states in
the two theories exactly match.
The nine-dimensional heterotic coupling
is  $\lambda_H^{(9)}=\lambda_H^{(10)}/ \sqrt{R_H}$ and similarly for
$\lambda_I^{(9)}$.  We thus have the following bounds in Phase I:
\beq
    {{{(R^{\ast})}^2\over R_H^{5/2}}~<~\lambda_I^{(9)}~<~{1\over
        {(R^{\ast})}^{1/2}\lambda_H^{(10)}}}
\label{rego}
\eeq
and
\beq
        \lambda_H^{(9)}~<~{R_H^{3/2}\over(R^\ast)^2}~.
\label{lamnin}
\eeq

If we fix $\lambda_H^{(10)}$ at a sufficiently small value,
our calculation of the heterotic cosmological constant $\Lambda^\ninen$,
together with the arguments of Refs.~\cite{nine,ten},
suggest that for $R_H>0$ the heterotic theory should flow in the direction of
increasing $R_H$ (or decreasing $\lambda_H^{(9)}$) towards the supersymmetric
$SO(32)$ endpoint.  There is a subtlety in the analysis as $R_H\rightarrow
\infty$ \cite{ten}, but this conclusion remains valid \cite{BD}.
The soliton should behave fundamentally in this phase of the theory,
but is clearly unstable.
For $\lambda_H^{(10)}$ large enough but still in Phase~I, our perturbative
calculation of the cosmological constant of the dual theory indicates
that the dual theory also flows to the supersymmetric
$SO(32)$ endpoint.  Note that the $R_H=0$ endpoint is not in the
Phase~I region for any value of the coupling $\lambda_H^\tenn$.

Since there is a phase transition connecting Phase~I to Phase~II, we
expect a different behavior in Phase~II.  We cannot analyze the dual
theory in Phase~II because of the presence of tachyons,
but we might try to relate the heterotic theory
to M-theory.  This makes sense since the length $\rho$ of the eleventh
dimension, defined in M-theory units as
\beq
        \rho ~\equiv~{{[\lambda_H^{(10)}]}^{2/3}\over R_H^{2/3}}~,
\label{M}
\eeq
satisfies
\beq
      \rho ~>~ { [\lambda_H^{(10)}]^{1/3}\over (R^{\ast})^{2/3}}~.
\label{Mtwo}
\eeq
Thus $\rho$ is bounded from below.
The low-energy analysis of Ref.~\cite{hor} seems to indicate
that there are no stable solutions of M-theory on a line segment
breaking supersymmetry in nine dimensions.  This,
combined with the fact that there is a phase transition separating Phase~I
from Phase~II, leads to several conjectures for the behavior of
our model.  Since it turns out \cite{BD} that both the ten-dimensional
$SO(32)$ and $E_8\times E_8$ supersymmetric heterotic strings
are continuously connected via interpolating models to
the $SO(16)\times SO(16)$ theory, one conjecture is that the theory flows to a
strongly
coupled $E_8\times E_8$ theory, or equivalently to  M-theory on $S^1/{\IZ_2}$.
  From what we have said about the expected behavior of the soliton at weak
coupling,
it is plausible that there is a phase transition at $R_H=0$ such that
when the $T$-dual ten-dimensional theory has nonzero coupling
$\lambda_H'\equiv \lambda_H^{(10)}\sqrt{\alpha'}/R_H$,
a fundamental $SO(16)\times SO(16)$  string in ten
dimensions behaves as an $E_8\times E_8$ string.
The subtlety discussed in Ref.~\cite{ten} further indicates that the boundary
of the
Phase~I region at $\lambda_H^{(10)}=R_H^2/{(R^{\ast})}^2$ is stable against
falling into Phase~I, at least for small $R_H$.
Moreover, by continuity arguments, we expect that this result can be
extended for all $R_H$.  However, the
above argument does not apply in Phase~II, and thus we have the interesting
possibility that this theory might spontaneously compactify sufficiently many
additional dimensions so that
it could be stabilized, for example, via a gluino condensate as in the model
of Ref.~\cite{hor}.

\begin{figure}[tb]
\centerline{
         \epsfxsize 4.0 truein \epsfbox {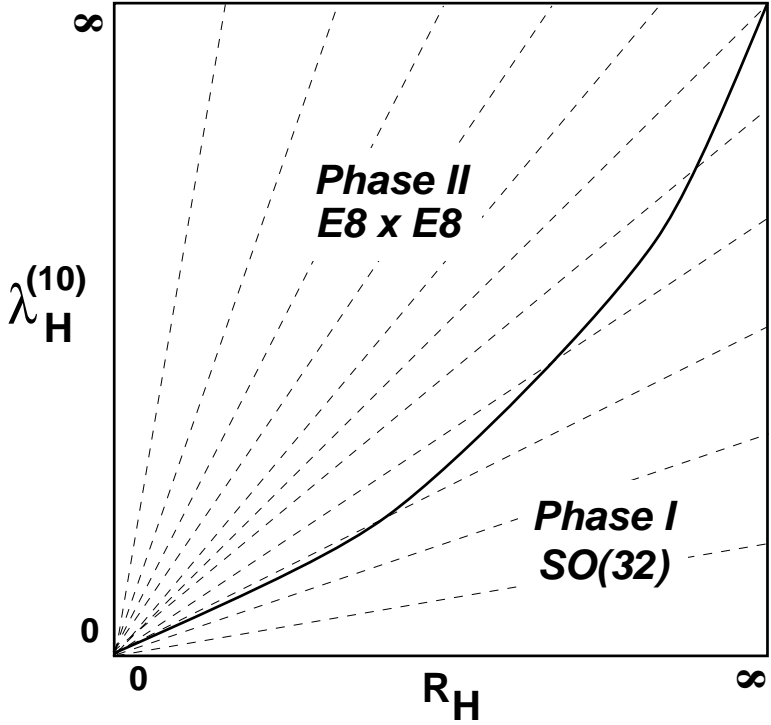}
  }
\caption{
     The proposed phase diagram for the \protect\sosixteen\ interpolating
     model.  For $\lambda_H^\tenn < R_H^2 / (R^\ast)^2$, the theory
     is in Phase~I and flows to the ten-dimensional weakly coupled heterotic
     supersymmetric $SO(32)$ theory.
     For $\lambda_H^\tenn > R_H^2 / (R^\ast)^2$, by contrast,
      the theory is expected to flow to the ten-dimensional strongly
     coupled supersymmetric $E_8\times E_8$ string.  On the boundary, the
     theory is stable against passing into Phase~I.  The dotted lines
     are contours of constant
     $T$-dual coupling $\lambda_H^\prime \equiv
\sqrt{\alpha'}\lambda_H^\tenn/R_H$.
    }
\label{phases}
\end{figure}

The results of our stability analysis can thus be summarized
as in Fig.~\ref{phases},
which shows our proposed phase diagram for the \sosixteen\ interpolating model.

\section{Conclusions}
\setcounter{footnote}{0}

We have constructed an exact strong/weak coupling dual for a
nonsupersymmetric, tachyon-free, heterotic theory that interpolates between
the ten-dimensional supersymmetric $SO(32)$ theory and the ten-dimensional
nonsupersymmetric $SO(16)\times SO(16)$ theory.  Our duality relation is
valid within a range of our interpolating parameter
corresponding to the tachyon-free phase of the dual interpolating theory.
In this tachyon-free phase, the two dual theories behave identically, have the
same
massless states, and flow to the
supersymmetric $SO(32)$ theory.  Furthermore, the dual theory contains a
soliton
that should behave as a fundamental $SO(16)\times SO(16)$ string at
sufficiently strong dual coupling.
We have conjectured that outside of this range, the strongly coupled
heterotic theory --- including the ten dimensional $SO(16)\times SO(16)$ theory
itself ---
flows to the supersymmetric strongly coupled $E_8\times E_8$ theory or
to some compactification of M-theory in which supersymmetry breaking occurs
in a stable way.

\bigskip
\medskip
\leftline{\large\bf Acknowledgments}
\medskip

We are happy to thank K.S.~Babu, K.~Intriligator, J.~March-Russell,
R.~Myers, S.~Sethi, F.~Wilczek, E.~Witten, and especially
A.~Sagnotti for discussions.
This work was supported in part by
NSF Grant No.\  PHY-95-13835
and DOE Grant No.\ DE-FG-0290ER40542.

\bigskip
\medskip

\bibliographystyle{unsrt}

\end{document}